\documentclass[12pt]{article}
\usepackage{amsfonts}
\usepackage{amsmath}
\usepackage{epsfig}

\newcommand{\beq}{\begin{equation}}
\newcommand{\eeq}{\end{equation}}
\def\tro{t_0}
\def\simlt{\stackrel{<}{{}_\sim}}

\begin{document}

\title{\hspace{4.1in}{\small CERN-PH-TH/2006-099} 
\\
\vspace*{1cm} The Price of WMAP Inflation in Supergravity}
\author{\vspace*{1.0cm} J. Ellis$^{a}$, Z. Lalak$^{b}$, S. Pokorski$^{b}$ and K. Turzy\'nski$^{b}$ 
 \\
$^{a}$ Theory Division, CERN, 1211 Geneva 23, Switzerland\\
$^{b}$ Institute of Theoretical Physics, University of Warsaw,\\ 
       00-681 Warsaw, Poland}
\date{}
\maketitle
\vspace*{2ex}
\begin{abstract}
\large
The three-year data from WMAP are in stunning agreement with the simplest possible
quadratic potential for chaotic inflation, as well as with new or symmetry-breaking inflation.
We investigate the possibilities for incorporating these potentials within supergravity,
particularly of the no-scale type that is motivated by string theory. Models with inflation
driven by the matter sector may be constructed in no-scale supergravity, if the moduli are 
assumed to be stabilised by some higher-scale dynamics and at the expense of some fine-tuning.
We discuss specific scenarios for stabilising the moduli {\it via} either D- or F-terms in
the effective potential, and survey possible inflationary models in the presence of
D-term stabilisation.
\end{abstract}
\thispagestyle{empty}

\newpage

\section{Introduction}

The dominant paradigm for explaining the great size, age and flatness of the Universe,
as well as the absence of unwanted primordial relics such as magnetic monopoles or 
gravitinos, is cosmological inflation. According to this hypothesis, very early in its
history, the Universe underwent an epoch of near-exponential expansion, dubbed
cosmological inflation. During this period, the energy density of the Universe is
hypothesized to have been dominated by a near-constant vacuum energy density.
Within the general context of quantum field theory, the favoured origin of this vacuum
energy density is the potential energy $V$ of some elementary scalar field $\phi$,
called the inflaton. Consistency with the magnitude of primordial density perturbations
inferred from the pioneering observations by the COBE satellite, as well as subsequent
experiments, suggests that this potential energy density $V \ll m_P^4$. 

In order for the
inflationary epoch to last long enough, and for the density perturbations to remain small,
the inflationary potential must obey certain slow-roll conditions:
\begin{equation}
\epsilon \; \equiv \; \frac{m_P^2}{2}\left( \frac{V^\prime}{V}\right)^2, \;
\eta \; \equiv \; m_P^2 \left(\frac{V^{\prime\prime}}{V}\right) \;
 \ll \; 1.
\label{slowroll}
\end{equation}
In principle, both the slow-roll parameters $\epsilon$ and $\eta$ may be extracted
from the spectra of density perturbations, in particular {\it via} the scalar spectral index
$n_s = 1 - 6\epsilon + 2\eta$ and the amplitude ratio of tensor and scalar perturbations
$r = 16\epsilon$. Also observable in principle is the running of the scalar spectral index,
$d n_s / d {\rm ln} k$, but this is expected to be very small in slow-rolling inflationary
models. 

\begin{figure}
\hspace{-1.0cm}
\includegraphics*[height=7cm]{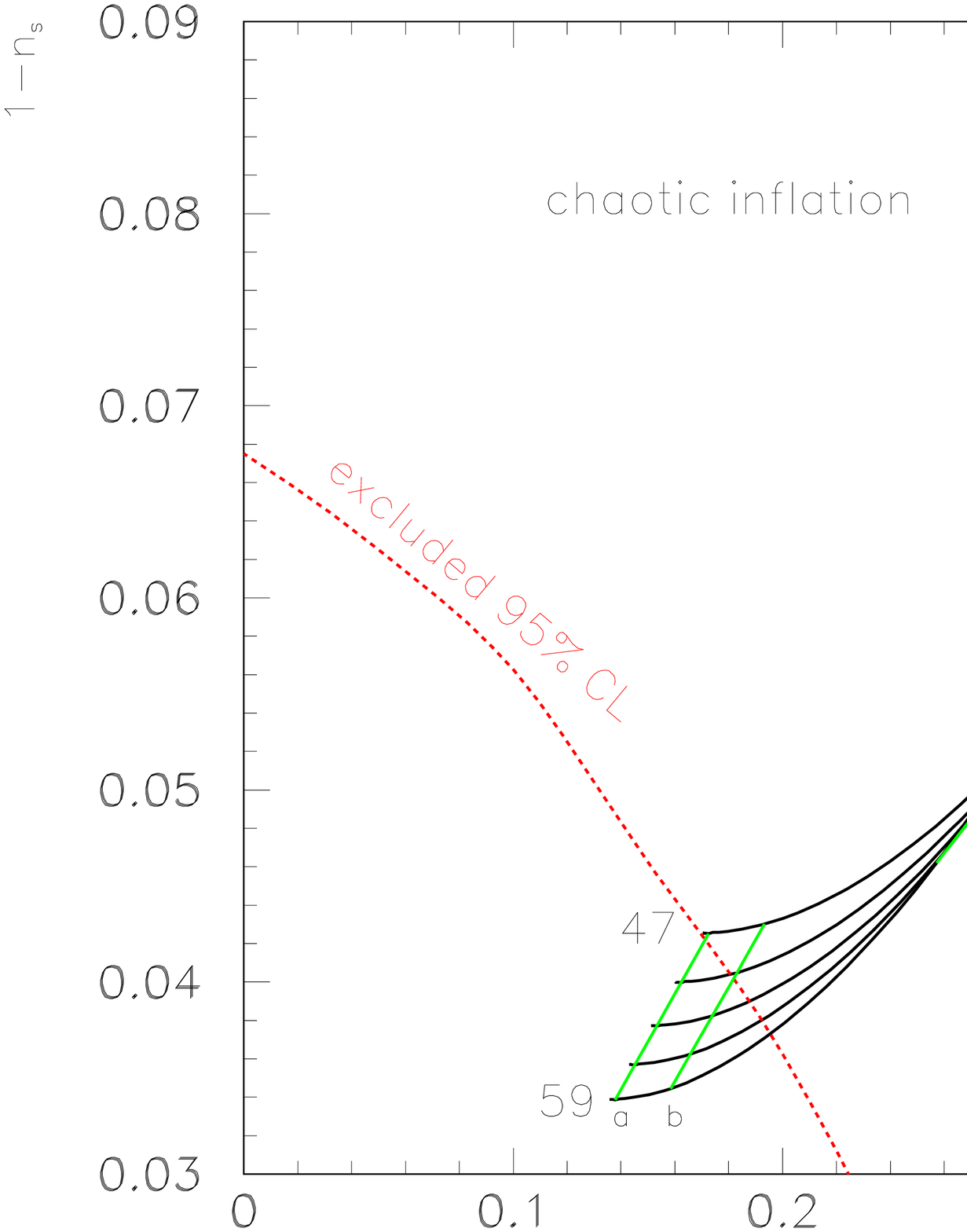}
\hspace{0.5cm}
\includegraphics*[height=7cm]{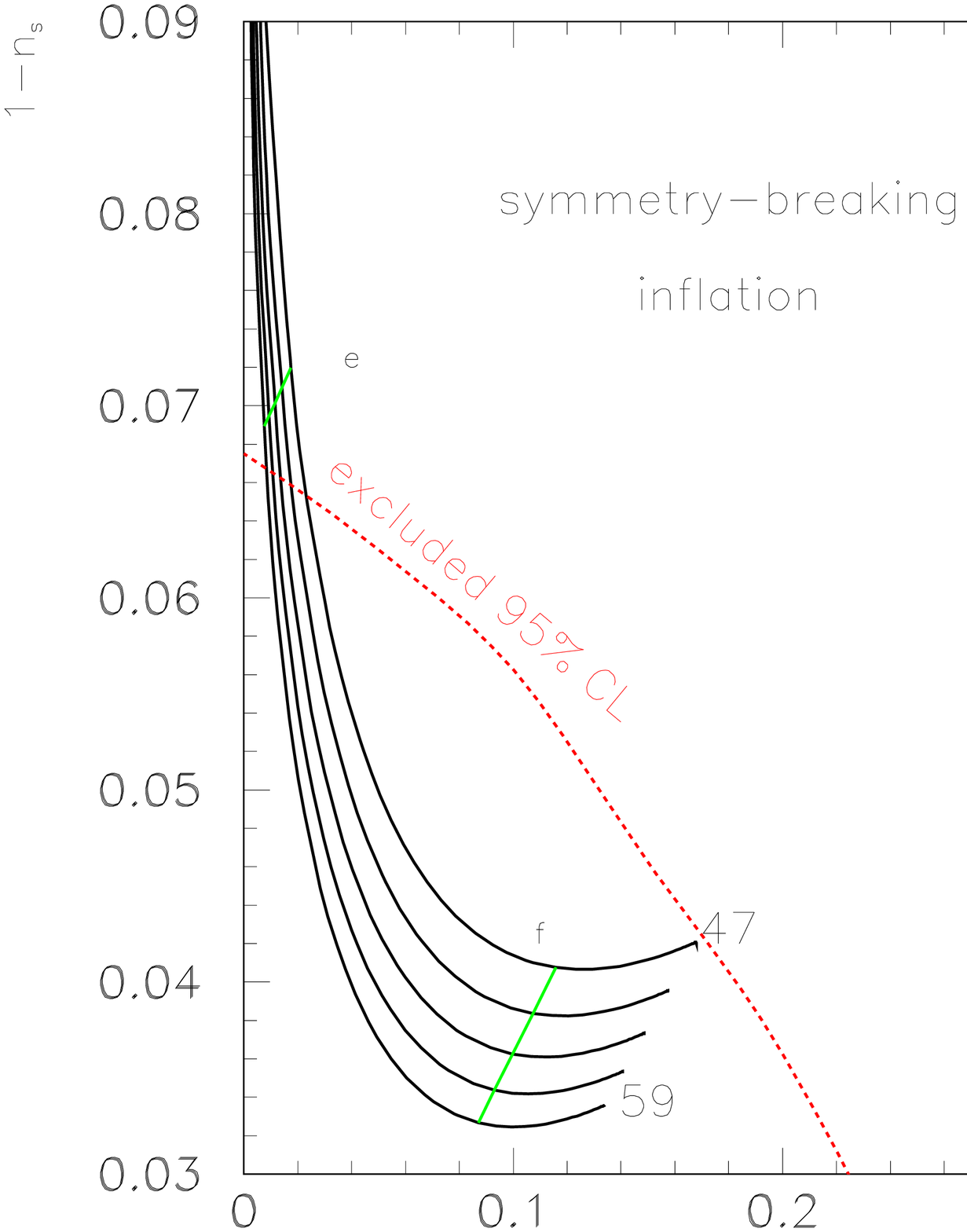}
\hspace{0.5cm}
\includegraphics*[height=7cm]{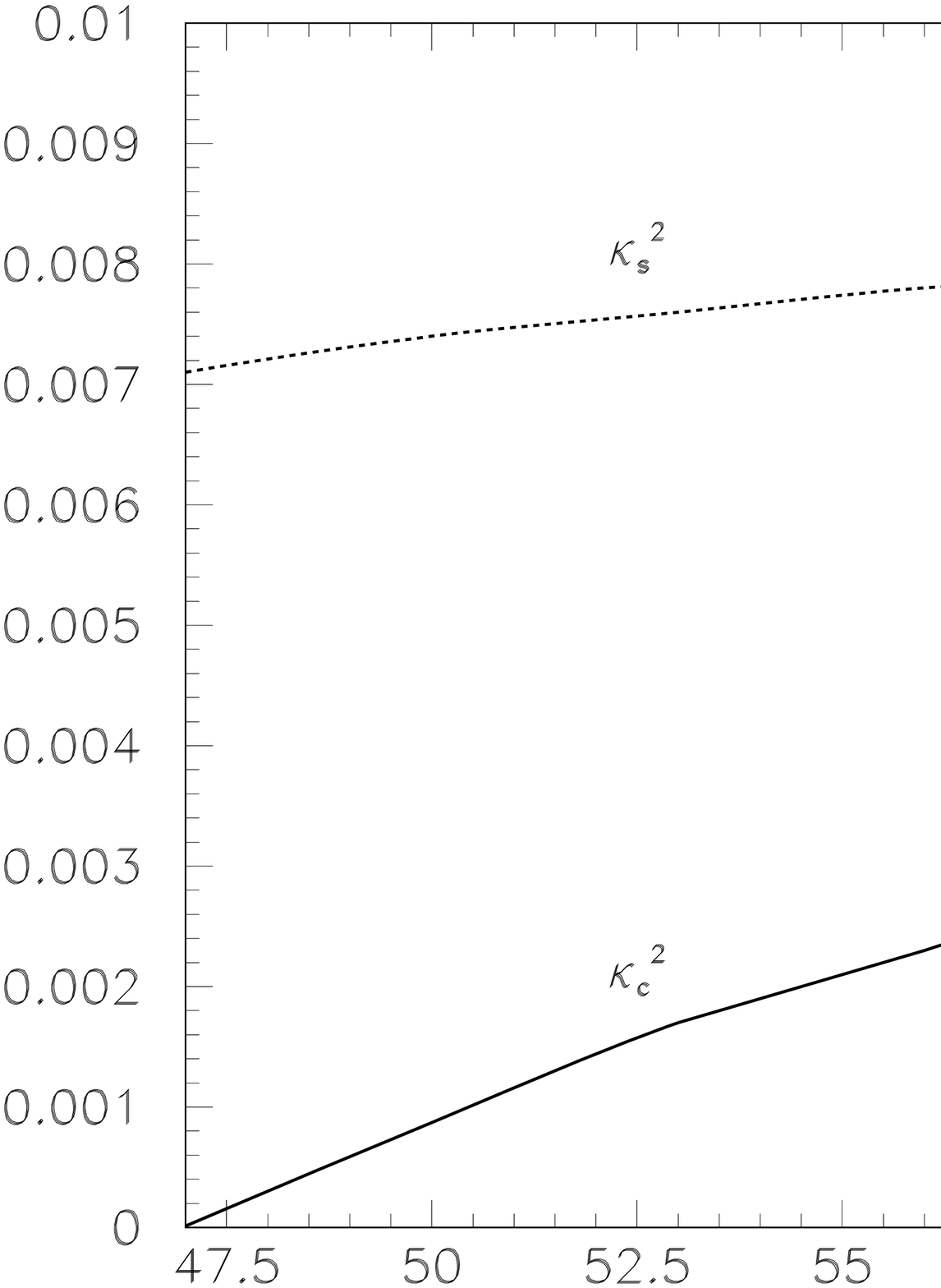}
\caption{\it Confrontation of low-order polynomial inflationary potentials
with the three-year data from WMAP~\protect\cite{Spergel:2006hy}, 
in combination with other data~\protect\cite{Seljak:2006bg}.
The deviation from scale invariance of the scalar spectral index,
$1-n_s$, is shown as a function of the tensor-to-scalar ratio $r$ for chaotic
inflation (left) and new inflation (centre). The lines correspond to 
$N=47,50,53,56,59$ e-foldings, as indicated. The green (light gray) lines denoted by
$\mathsf{a}-\mathsf{d}$ correspond to
$\kappa_c^2=10^{-4},10^{-3},10^{-2},10^{-1}$ in (\ref{pro:pot1}), and the
green (light gray) lines denoted by $\mathsf{e},\mathsf{f}$ correspond to
$\kappa_s^2=10^{-2},10^{-3}$ in (\ref{pro:pot2}).  The dashed curves show the
regions of the parameters allowed by WMAP3 and other data~\protect\cite{Seljak:2006bg}
at the 95\% CL. Right panel
displays the corresponding 95\% upper limits on $\kappa_c^2$ and $\kappa_s^2$
as functions of the numbers of inflationary e-foldings in the chaotic and
new inflationary models, respectively.
\label{pro:fig}}
\end{figure}

Important constraints on the slow-roll parameters have been obtained in the past,
using the first-year data from WMAP and other observational inputs.
Qualitatively new insight has recently been provided by the new three-year WMAP 
data (WMAP3), which are strikingly consistent with simple
models of inflation based on an elementary scalar inflaton field whose
potential is just a low-order polynomial~\cite{Spergel:2006hy} 
{}\footnote{Note that this statement is also supported by the WMAP3 data 
analysis given in~\cite{Kinney:2006qm}, which yields a larger allowed 
region.}. 
In particular, the data
are consistent with the simplest model for chaotic inflation with a
quadratic potential $V(\phi)=\frac{1}{2}m^2\phi^2$, as seen in Fig.~\ref{pro:fig}.
They also constrain severely possible modifications, e.g., of the form:
\beq
\label{pro:pot1}
V(\phi) = \frac{1}{2}m^2\phi^2\left(1+\frac{1}{2}\kappa_c^2\phi^2\right) .
\eeq
As seen in the right panel of Fig.~\ref{pro:fig}, the supplementary parameter $\kappa_c$ 
cannot exceed a few per mille, with the exact number
depending of the number of e-foldings during the inflationary epoch. As seen in
Fig.~\ref{pro:fig}, the WMAP3 data~\cite{Spergel:2006hy} are also consistent with 
another very simple potential:
\beq
\label{pro:pot2}
V(\phi) = \lambda m_P^4 \left( 1-\kappa_s^2\frac{\phi^2}{m_P^2}\right)^2 ,
\eeq
with the requirement, seen also in the right panel of Fig.~\ref{pro:fig}, 
that $\kappa_s^2$ cannot exceed one
per cent if inflation is to start with the small initial value of the
inflaton field~\footnote{Note, however, that this potential also allows
inflation with a large value of the inflaton field.}, i.e., on the branch
$|\phi|\leq m_P/\kappa_s$. This scenario is known as new or
symmetry-breaking inflation. These and other models make characteristic
predictions for the spectral index of scalar density perturbations, $n_s$,
and for the tensor-to-scalar ratio $r$. The left and central panels of Fig.~\ref{pro:fig}
display the constraints on
$n_s$ and $r$ inferred from a combination of WMAP3~\cite{Spergel:2006hy} with other data~\cite{Seljak:2006bg}.
These are confronted with the different
predictions of the potentials given in (\ref{pro:pot1}) and
(\ref{pro:pot2}), assuming negligible running of the scalar spectral
index, as predicted in these models.

It has long been a challenge to embed inflation in realistic models of
particle physics. There are two aspects to this problem: identifying a
suitable candidate for the inflaton, and ensuring that it has a suitable
effective potential. We take the point of view that realistic particle
models should incorporate supersymmetry~\cite{ENOT}, which incidentally provides many
scalar fields that one might hope to exploit as an inflaton. None of the
MSSM fields fits the bill, but the scalar partner of one of the heavy
singlet neutrinos in a seesaw model of neutrino masses might be suitable~\cite{MSYY}.
In the framework of global supersymmetry, such a sneutrino would naturally
possess a simple quadratic potential without significant perturbative
corrections, and its mass would fit naturally with the estimate of $m \sim
2 \times 10^{13}$~GeV required for the simplest chaotic inflation model
(\ref{pro:pot1}).

However, a more appropriate framework for describing inflation is
presumably that offered by local supersymmetry. The low-energy limit of
any supersymmetric theory of quantum gravity is necessarily some such
supergravity theory. Moreover, since inflation is a fundamental property
of the evolution of space-time, it must involve the gravitational sector
of the microscopic theory, in which case supergravity is likely to play an
essential role. The support of the data for the simplest inflationary
models is then highly non-trivial information, since the effective
potentials yielded by supergravity theories generically have forms that
are more complicated than (\ref{pro:pot1}) or (\ref{pro:pot2}). Indeed,
many supergravity potentials have minima with vacuum energy densities that
are negative and ${\cal O}(m_P^4)$.

The general term for this troublesome feature of supergravity is the
$\eta$ problem, namely the statement that supergravity with simple forms
for the K\"ahler potential for scalar superfields, such as minimal
supergravity, naturally predict large mass terms for all the scalars,
destroying the flatness of any candidate inflationary direction in the
effective potential. One possible way out of this problem is to postulate
a K\"ahler potential whose prefactor $e^K$ dependens on the scalar fields
in a manner softer than an exponential. The non-minimal form of the
K\"ahler potential which is particularly well motivated from the point of
view of string theory is the no-scale structure that appears naturally in
compactifications of higher-dimensional superstring models. The great
advantage of the no-scale structure is that it leads to a semi-positive
definite scalar potential, very similar to a globally supersymmetric one~\cite{Ellis:1983sf}.

Many superstring theories contain moduli fields $T_i$ whose effective potentials
are flat classically. These no-scale models are characterized by
effective low-energy supergravity theories containing terms in the
K\"ahler potential $K$ of the characteristic logarithmic form:
$K \ni - \log(T_i + \bar{T}_i)$. Since the effective potentials of these
moduli fields vanish classically, one might expect them to remain small even
when corrections are calculated, so that they might be candidates for the inflaton field.
Efforts to implement modular inflation continue and provide interesting
and encouraging results.
However, here we prefer to explore the other possibilities offered by the
matter sectors of string theories, which are typically rather rich in the
structures of their superpotentials and K\"ahler potentials, potentially
offering ample opportunities to generate inflationary behaviour.

Specifically, in this paper we examine in some detail conditions necessary for 
matter-driven inflation to take place in a locally supersymmetric model. For simplicity, 
we consider here a no-scale model with a single modulus $T$ and a matter field $\Phi$,
with a no-scale K\"ahler potential of the form:
\begin{equation}
K \; = \; - 3 \log (T + \bar{T} - 2 | \Phi |^2)
\label{simpleK}
\end{equation}
and a matter superpotential to be discussed below.
We find that the matter sector in such a simplified no-scale supergravity may
indeed lead to acceptable inflation, provided that the $T$ modulus is stabilised
appropriately. One must also apply specific fine-tuning conditions
on the inflationary superpotential, which depend on the precise form
of the constraint on the modulus $T$. In at least some such models, the
naturalness of these fine-tuning requirements is not obvious from the
point of view of the effective low-energy theory.

We then go on to consider explicit models for the stabilization of the
modulus $T$. One is D-type stabilisation made
possible by supersymmetric D-terms, and the other is F-type stabilisation, where the
superpotential depends on a $T$ modulus.  
As we show below, model building is much easier in the case of D-stabilisation, as
the scales in the sector of the matter-type inflaton $\Phi$ and in the
$T$-modulus sector can decouple from each other.

\section{No-scale Structure and Inflation from Matter Fields}
\label{sec:Tfixed}

As already mentioned, minimal supergravity models with canonical
kinetic terms for the matter fields have a serious disadvantage when it
comes to inflationary scenarios. The trouble is known as the $\eta$
problem, and is due to the common prefactor $e^K = e^{|\Phi|^2}$ in the
scalar potential, where we denote by $\Phi$ the matter-type
inflaton~\footnote{Here and subsequently, we use the system of units with
$m_P=1$.}. However, it is possible to modify the K\"ahler potential,~\cite{Ellis:1984bf},
 so as
to obtain simple models that are very similar to those of the globally
supersymmetric case~\footnote{The $\eta$ problem itself can also be cured
in other ways, for instance by using a phase of a field with canonical
K\"ahler potential, and a K\"ahler potential of the form $K=1/2 (M +
\bar{M})^2$ which makes the factor $e^K$ insensitive to the imaginary part
of $M$~\cite{yanagida}. Nevertheless, the models closest in their forms to
globally supersymmetric Lagrangians are offered by the no-scale K\"ahler
potentials.}.
These are
no-scale models of the type that arises naturally in compactifications of
string models~\cite{Witten:1985xb},~\cite{Derendinger:1985cv},~\cite{Derendinger:2005ph},
 in which case
\begin{equation}
K=- n_1 \log (T_1 + \bar{T}_1
- 2 |\Phi_1|^2) - n_2 \log (T_2 + \bar{T}_2 - 2 |\Phi_2|^2) - n_3 \log
(T_3 + \bar{T}_3 - 2 |\Phi_3|^2),
\label{noscale}
\end{equation}
where $n_1 + n_2 + n_3 =3$ and $n_i \geq 0\;$ for $i = 1, 2, 3$.

{}As already mentioned in the Introduction, for simplicity, we consider the minimal case  
$K=-3 \log (T + \bar{T} - 2 |\Phi|^2)$, for which the kinetic part of the Lagrangian
for the fields $T$ and $\Phi$ reads:
\beq
\label{fix:kin1}
\mathcal{L}_\mathrm{kin} = \frac{1}{(T+\bar{T}-2|\Phi|^2)^2}\left(\partial_\mu\bar{T},\partial_\mu\bar{\Phi}\right) 
\left(\begin{array}{cc} 3 & -6\bar{\Phi} \\ -6\Phi & 6(T+\bar{T}) \end{array}\right)
\left(\begin{array}{c}\partial^\mu T \\ \partial^\mu\Phi 
\end{array}\right).
\eeq
This expression can easily be written in an equivalent form:
\begin{eqnarray}
\mathcal{L}_\mathrm{kin} &=& \frac{1}{2} \left[\partial_\mu\sqrt{\frac{3}{2}}\ln 2(t-|\Phi|^2)\right]^2 + \frac{3\left[\partial_\mu t'+\mathrm{i}(\bar{\Phi}\partial_\mu \Phi-\Phi\partial_\mu \bar{\Phi})\right]^2}{4(t-|\Phi|^2)^2} 
\label{fix:kin2} 
+\frac{3\partial_\mu\bar{\Phi} \partial^\mu \Phi}{(t-|\Phi|^2)}  ,
\end{eqnarray}
where $t \equiv \mathrm{Re}\,T$, $t' \equiv \mathrm{Im}\,T$. Note, that the above kinetic term is positive definite as long as $|\Phi| < \sqrt{t}$. 
Assuming an arbitary superpotential~\footnote{Addressing the possible 
origins of non-trilinear terms in the effective 
superpotential goes beyond the scope of this paper, but we note that various mechanisms for generating them may exist, see for instance 
\cite{Giudice:1988yz}.} $W(\Phi)$ which does not depend on $T$,  
one obtains the scalar potential 
\begin{equation} 
\label{nosc}
V_F(\Phi)= \frac{|\partial_\Phi W|^2}{6 (T + \bar{T} - 2 |\Phi|^2)^2}.
\end{equation}
In a complete model, $T$ is a dynamical degree of freedom, and one can
readily convince oneself that in fact it is necessary to stabilise T. To
see this, let us compute the $\eta$ and $\epsilon$ parameters along the
direction of the real part of $T$:
\begin{equation} 
\eta_T = \frac{1}{2 V} \frac{\partial }{\partial t} \left (
(K^{-1})^{\bar{T}T} \frac{\partial V}{\partial t} \right ), \qquad
\epsilon_T = \frac{1}{4} (K^{-1})^{\bar{T}T} \left ( \frac{1}{V}
\frac{\partial V}{\partial t} \right )^2 .
\end{equation}
Because of the particular structure of the scalar
potential (\ref{nosc}), where the dependence on the real part of $T$ comes
only from the denominator, one finds, independently of the form of $W$, the
results $\eta_T = \epsilon_T = 4/3$.  This means that the curvature of
the potential, and hence the driving force acting on $T$ along the 
direction of ${\rm Re}\, T$, is large everywhere, making it impossible 
to generate any inflation in the $\Phi$ sector.

We therefore discuss inflation from matter fields, assuming that the 
only modulus $T$ in this simple model is frozen by some, as yet 
unspecified, mechanism~\footnote{We note that fluxes and non-perturbative 
gaugino condensates may determine many or all moduli in semi-realistic models.}. Then 
the part of the supergravity Lagrangian involving the $\Phi$ field can be 
obtained by combining (\ref{fix:kin1}) with $T=\mathrm{const}$ and (\ref{nosc}):
\beq
\label{fix:lag}
\mathcal{L}_\Phi=\frac{6(T+\bar{T})}{(T+\bar{T}-2|\Phi|^2)^2} \partial_\mu
\Phi \partial^\mu \bar{\Phi} - V_F(\Phi) \qquad.
\eeq
We can now rewrite the kinetic term for the modulus $|\Phi|$ coming from  (\ref{fix:lag}) in terms of the canonically normalised
field $\phi$ (we freeze the phase of $\Phi$ for the moment):
$|\Phi|/\sqrt{t}=\mathrm{tanh}(\phi/\sqrt{6})$. The effective potential $V_F(\Phi)$
may therefore be written in the form
\begin{equation}
V_F(\phi) \; = \; \frac{1}{24 t^2} \left[|\partial_\Phi W|^2\right]_{|\Phi| = \sqrt{t}\mathrm{tanh}(\phi/\sqrt{6})}
\times \mathrm{cosh}^4(\phi/\sqrt{6}).
\label{generalpot}
\end{equation}
Assuming the simplest forms
\begin{equation}
W(\Phi)=\frac{1}{2}M\Phi^2 \; {\rm or} \; 
W(\Phi)=\frac{1}{3}\lambda\Phi^3
\label{or}
\end{equation}
for the superpotential, corresponding to
\begin{equation}
\left[|\partial_\Phi W|^2\right]_{|\Phi| = \sqrt{t}\mathrm{tanh}(\phi/\sqrt{6})} = M^2 
\mathrm{tanh}^2(\phi/\sqrt{6}) t \; {\rm or} \; \left[|\partial_\Phi 
W|^2\right]_{|\Phi| = \sqrt{t}\mathrm{tanh}(\phi/\sqrt{6})} = \lambda^2\mathrm{tanh}^4(\phi/\sqrt{6}) t^2,
\label{or2}
\end{equation} 
we find the resulting scalar potential
\beq
V_F(\phi) = 
\frac{M^2}{96t}\mathrm{sinh}^2\left(\sqrt{\frac{2}{3}}\phi\right)\qquad\textrm{or}\qquad V_F(\phi) = 
\frac{\lambda^2}{24}\mathrm{sinh}^4\left(\sqrt{\frac{1}{6}}\phi\right),
\eeq
respectively, which do not satisfy the slow-roll conditions. 

Hence, with $T$ stabilised one should construct models which go beyond the quadratic 
superpotential for $\Phi$. From the point of view of an effective
low-energy theory with a polynomial superpotential in $\Phi$, obtaining an inflationary potential 
that is a simple quadratic in the canonically-normalized inflaton field $\phi$, as apparently preferred by the WMAP3 data, would seem quite unnatural.
Nevertheless, a potential that is purely quadratic in $\phi$ may in principle be achieved by postulating a non-polynomial superpotential $W(\Phi)$,
which gives 
\begin{equation}
\frac{\partial W}{\partial \Phi} \; \propto \; {\rm artanh} \left( \frac{|\Phi|}{\sqrt{t}} \right) \times
\left(1 - \frac{|\Phi|^2}{t}\right).
\label{wackyW}
\end{equation}
However, this form looks all the more fine-tuned because the value of $t$ is to be fixed dynamically by some unknown mechanism.

Another simple choice would be
\beq
\label{fix:super2}
W=W_0+\mu^2\Phi+\frac{1}{3}d\Phi^3\; , 
\eeq
from which one obtains the scalar potential 
\beq
\label{fix:new1}
V_F(\Phi) = \frac{\mu^4}{24(t-|\Phi|^2)^2}\left[\left(1- 
\frac{|\Phi|^2}{\Phi_0^2}\right)^2+\frac{4|\Phi|^2}{\Phi_0^2}\cos^2 
\theta\right] ,
\eeq
where $\theta \equiv \mathrm{arg}\,\Phi$ and $\Phi_0^2=\mu^2/d$.
For $|\Phi|^2<t,\Phi_0^2$, the contributions to the mass of $|\Phi|$ coming 
from the numerator and the denominator of (\ref{fix:new1}) can 
approximately cancel each other,
making this degree of freedom much lighter than $\theta$ and hence a good 
candidate for the inflaton field. It is, therefore, reasonable to assume 
that $\theta$ takes the value at its minimum, namely $\pi/2$, already  at 
the onset of inflation, and the potential (\ref{fix:new1}) can then be 
effectively rewritten in terms of $|\Phi|$ only. For $|\Phi|\ll t$, one 
can expand the potential (\ref{fix:new1}) in $|\Phi|^2/t$, obtaining an 
effective potential of the type (\ref{pro:pot2})
with $\kappa_s^2=t/\Phi^2_0-1$. Hence, symmetry-breaking inflation is 
possible, provided that the parameters of the matter superpotential and the 
value $t$ of the fixed modulus are such that:
\beq
\label{fix:finetuning}
0<1-\frac{\Phi^2_0}{t} \simlt 10^{-2}.
\eeq
We note that the potential (\ref{fix:new1}) reduces to the effective 
potential (\ref{pro:pot2}) only in the limit $|\Phi|^2\ll t$ and, unlike 
(\ref{pro:pot2}), it does not allow for chaotic initial conditions,
as $\epsilon \geq 4/3$ and $\eta \geq 4/3$ for $|\Phi|>\Phi_0$.

Alternatively, instead of $T$ stabilisation one could 
postulate stabilisation of the combination $t-|\Phi|^2$, which appears 
in the K\"ahler potential. In this case, the canonically-normalized inflaton field $\phi$ is given by
\begin{equation}
\phi \; = \; \frac{\sqrt{6}|\Phi|}{\sqrt{t-|\Phi|^2}}
\label{otherfix}
\end{equation}
and the effective inflationary potential is then simply proportional to $|\partial W/\partial \Phi |^2$.
Contrarily to the previous case, if $W=\frac{1}{2}M\Phi^2$ the effective potential 
reduces to the simplest quadratic form allowed by WMAP3:
\begin{equation}
\label{fix:vother}
V_F = \frac{1}{2}m^2\phi^2 \qquad \textrm{where} \qquad m^2=\frac{M^2}{72(t-|\Phi|^2)}=\mathrm{const}.
\end{equation}

In order to go beyond the fixed-$t$ or fixed-$(t-|\Phi|^2)$ approximation, one needs
to solve the equations of motion for all relevant fields. 
Since the kinetic terms are complicated functions of
the original fields $T$, $\Phi$ and their derivatives, and the potential is a 
very complicated function of the canonically normalised fields, we shall use 
the following degrees of freedom: $|\Phi|$,  $\theta$,  
$\tau=\ln(2t-2|\Phi|^2)$ and $\zeta$, where
$t'=\mathrm{Im}\,T$ and $\mathrm{d}\zeta = \mathrm{d}t'-2|\Phi|^2
\mathrm{d}\theta$. The potential is, in general, a function of $\tau$, 
$|\Phi|$ and $\theta$. In terms of these fields, the equations of motion 
read: 
\begin{eqnarray} 
\label{eom1}
\ddot{\tau}+3H\dot{\tau}+4\dot{\zeta}^2e^{-2\tau}+2|\dot{\Phi}|^2e^{-\tau}+
2|\Phi|^2\dot{\theta}^2 e^{-\tau} &=& -\frac{2}{3}\frac{\partial 
V}{\partial\tau} \\ 
\label{eom1a}
\ddot{\zeta}+3H\dot{\zeta}-2\dot{\zeta}\dot{\tau} &=& -\frac{e^{2\tau}}{6}
\left(\frac{\partial V}{\partial t'}-\frac{1}{2|\Phi|^2}
\frac{\partial V}{\partial\theta} \right) \\
|\ddot{\Phi}|+3H|\dot{\Phi}|-|\dot{\Phi}|\dot{\tau}+|\Phi|\dot{\theta}^2 &=&
-\frac{e^{\tau}}{12}\frac{\partial V}{\partial |\Phi|} \\ 
\ddot{\theta}+3H\dot{\theta}+2\frac{|\dot{\Phi}|}{|\Phi|}\dot{\theta}-
\dot{\theta}\dot{\tau} 
&=& -\frac{e^{\tau}}{12|\Phi|^2}\frac{\partial V}{\partial \theta},
\label{eom2}
\end{eqnarray}
where the dot denotes differentiation with respect to cosmic time and the 
Hubble parameter is:
\beq
\label{eomh}
H^2 = 
\frac{1}{4}\dot{\tau}+\dot{\zeta}^2e^{-2\tau}+2|\dot{\Phi}|^2e^{-\tau}+
2|\Phi|^2\dot{\theta}^2e^{-\tau}+\frac{1}{3}V .
\eeq
The equations of motion (\ref{eom1a}) and (\ref{eom2}) show that, for 
$\tau$ approximately fixed and for the potential depending only on 
$|\Phi|$, the kinetic energy associated with the quantity $\zeta$ 
parametrising the imaginary part of $T$ and with the phase $\theta$ 
is redshifted away with the expansion of the Universe, and it is 
legitimate to restrict one's attention to $\tau$ and $|\Phi|$ in the 
discussion of inflation. In our numerical analyses of the models presented
in the following sections, we integrate the full set of the equations of motion
(\ref{eom1})-(\ref{eomh}), assuming that the initial values of
$\theta$, $\dot{\theta}$ and $\dot{\zeta}$ vanish.
We neglect the isocurvature perturbations, generically present in the multi-field
inflationary models, assuming that we can choose the parameters of the potential
so that the curvature of the potential in the directions transverse to the classical
trajectory is much larger than that in the longitudinal 
direction~\footnote{We thank
D.~Langlois for drawing our attention to this point.}.

\section{Modulus Stabilisation through the D-terms}

We now describe the forms of the D-terms in various cases, and the ways in which they may stabilize the modulus~\cite{Ciesielski:2002fs}-\cite{Villadoro:2005yq}.   

We first consider gauging the imaginary shift of the modulus $T$. 
We assume that $K=K(T + \bar{T})$, and consider the imaginary shift $T \rightarrow T + i \delta /2 \Lambda$, which is generated by the 
Killing vector $P_T = i \delta/2$ where 
$\delta $ is real. The prepotential $D$ fulfills the Killing equation $K_{T\bar{T}} \bar{P}_T =  i \partial D/ \partial T$, 
and generates the scalar potential $\delta V = 1/2 g^2 D^2 $.  In the case of a pseudo-anomalous 
U(1)~\footnote{I.e., one for which the sum of the charges of all fields on which the symmetry is realised linearly does not vanish.}, the gauge coupling 
depends on the modulus $T$ as follows: $g^{-2}={\rm Re}(T)$. 
One finds, upon solving the Killing equation, that
$D=- \frac{\delta}{2} \frac{\partial K}{\partial \bar{T}} + \xi$, where $\xi$ is a genuine constant. 
If $\xi \neq 0$, the gauged symmetry acts 
on the gravitino and becomes a local U(1) R-symmetry. To simplify the discussion, we set $\xi=0$, unless stated otherwise. 
To be consistent with the global supersymmetry algebra, the superpotential must be invariant under any local symmetry which 
is not an R-symmetry. 
The invariant superpotential which is a function of the modulus $T$ alone must be a constant, which is insufficient to stabilize the modulus $T$.  

The solution is to introduce more charged fields:
\begin{equation}
\delta \phi = i \Lambda q_\phi \phi; \;\; P_\phi=i q_\phi \phi,
\end{equation}
in which case the general D-term becomes
\beq
D=- \frac{\delta}{2} \frac{\partial K}{\partial \bar{T}} - q_\phi  \frac{\partial K}{\partial \phi} \phi \ . 
\eeq

\noindent
in order to begin the discussion of D-term stabilisation,
we assume initially that the superpotential does not depend on the modulus $T$. Then the only possibility for stabilizing 
this modulus is to rely on the D-terms, which give positive contributions to the scalar potential, as seen in
\cite{Lalak:2005hr}. 
One obvious case is to make the gauge coupling $T$-dependent: $g^2 = 1/t$. However, if this is the only source of $T$ dependence,
there is no stabilization, as the resulting contribution to the scalar potential is monotonic in $t$: $V_D = 1/(2 t) D^2 $.
To have a better chance for stabilization, one needs to introduce $T$ inside the D-term. One option  is to make $T$ charged 
under the U(1). According to the discussion from the previous subsection, there are two options. First, the U(1) may be 
an R-symmetry, in which case there appears an additional constant term in the D-term: $D \rightarrow D + \xi$, which indeed aids in stabilization. However, the superpotential must then be charged under the 
U(1), which means that (a) $\Phi$ must be charged, 
or (b) one introduces yet another field, say $M$, which is charged and mixes with $\Phi$ in the superpotential via a term of the form $M^\alpha \Phi^\beta$, so as
to give the proper charge to terms containing $\Phi$. Both options are disastrous for $\Phi$ inflation, since either they restrict very strongly 
the allowed form of the superpotential for $\Phi$ [case (a)] or lead to complicated multi-inflaton models [case (b)]. Hence we give up on the R-symmetric models. 

Instead, we continue with a symmetry which leaves the superpotential invariant. 
However, in this case we need yet another $T$-dependent contribution to the D-term, so there must exist other fields charged 
under the U(1), which we denote by $X_i$.  Let the relevant terms in the K\"ahler potential be $ K(X_i, \bar{X}_i) = \frac{|X_i|^2}{(T + \bar{T})^{n_i}}$. The D-term contribution to the scalar potential then reads
\begin{equation}
\label{d:vd}
V_D = \frac{g^2 }{2} \left ( \frac{3 \delta }{T + \bar{T} - 2 |\Phi|^2} + \sum_i^{} Q_{X_i}   \frac{|X_i|^2}{(T + \bar{T})^{n_i}} \right )^2 \, .
\end{equation}
It is obvious that, given the right  assignment of charges ($\delta$ and $Q_{X_i}$), one can stabilize the real part of 
$T$ once the spectator fields $X_i$ assume non-zero expectation values, \cite{Lalak:2005hr}. This conclusion is independent 
of the possible $T$-dependence of $g^2$. Of course, the question 
then arises of the source of the v.e.v. of the $X_i$. 
One needs to assume that there exists a sector in the theory, independent of the $X_i$, which stabilizes 
the $X_i$ and decouples  above the scale of $\Phi$ inflation. This type of assumption is often made in inflationary model building, 
and does not constrain the discussion in a significant manner.   It may be viewed as an additional fine-tuning, which is 
in any case present in inflationary models.

An interesting question concerns the issue of supersymmetry breaking after inflation. As the vacuum energy vanishes in the post-inflationary 
minimum, all D-terms and F-terms must become zero at that minimum  with the exception of the $F^T${}\footnote{The $\langle F^T \rangle$ is not bound to 
vanish, since, as long as the no-scale structure is not disturbed, $K_{T \bar{T}} |F^T|^2 = 3 e^K |W|^2$ identically.}, whose expectation
value is $\langle F^T \rangle = m_{3/2} \langle T + \bar{T} - 2 |\Phi|^2 \rangle = e^{K/2} \langle W \rangle  \langle T + \bar{T} - 2 |\Phi|^2 \rangle$. 
In all models we consider, this could be zero only if the expectation value of the complete superpotential, $\langle W \rangle$,
 vanishes. This is generically not the case, 
so supersymmetry is broken in the post-inflationary minimum. However, the scale of $\langle W \rangle$ is set by the expectation values 
of the fields $\Phi$ and $X_i$, which in turn determine the height of the inflationary potential. The additional freedom which helps to match 
the post-inflationary 
gravitino mass and WMAP normalisation lies in the fact that the WMAP 
normalisation constrains the ratio of the height of the potential and the $\epsilon$ 
parameter, and, if the initial conditions are set by the saddle point, it is just the constraint on the precision of initial localization 
of the inflaton near the saddle 
point (where $\epsilon$ goes towards zero). To study this issue reliably one needs to specify the model in more details than are  necessary to study the inflationary epoch alone, 
hence we do not pursue this discussion any further in the present paper. 

\section{Inflation in the Presence of D-term Stabilisation}

Assuming that we are able to stabilise the modulus $T$ with the help of a D-term, we have a number of options to create an inflationary epoch. The available scenarios can be grouped into three broad classes. 

\begin{figure}[t]
\begin{center}
\includegraphics*[height=7cm]{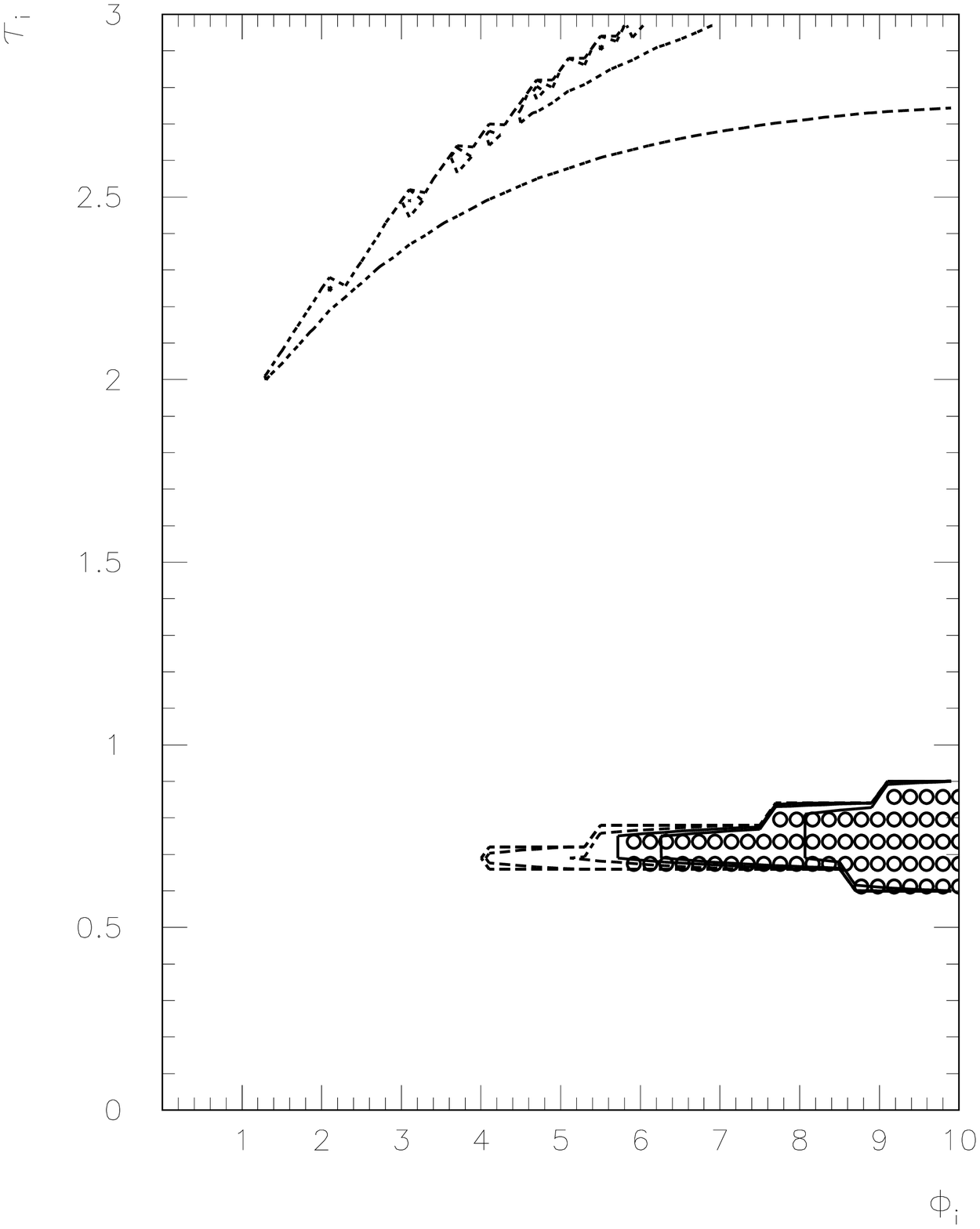}
\includegraphics*[height=7cm]{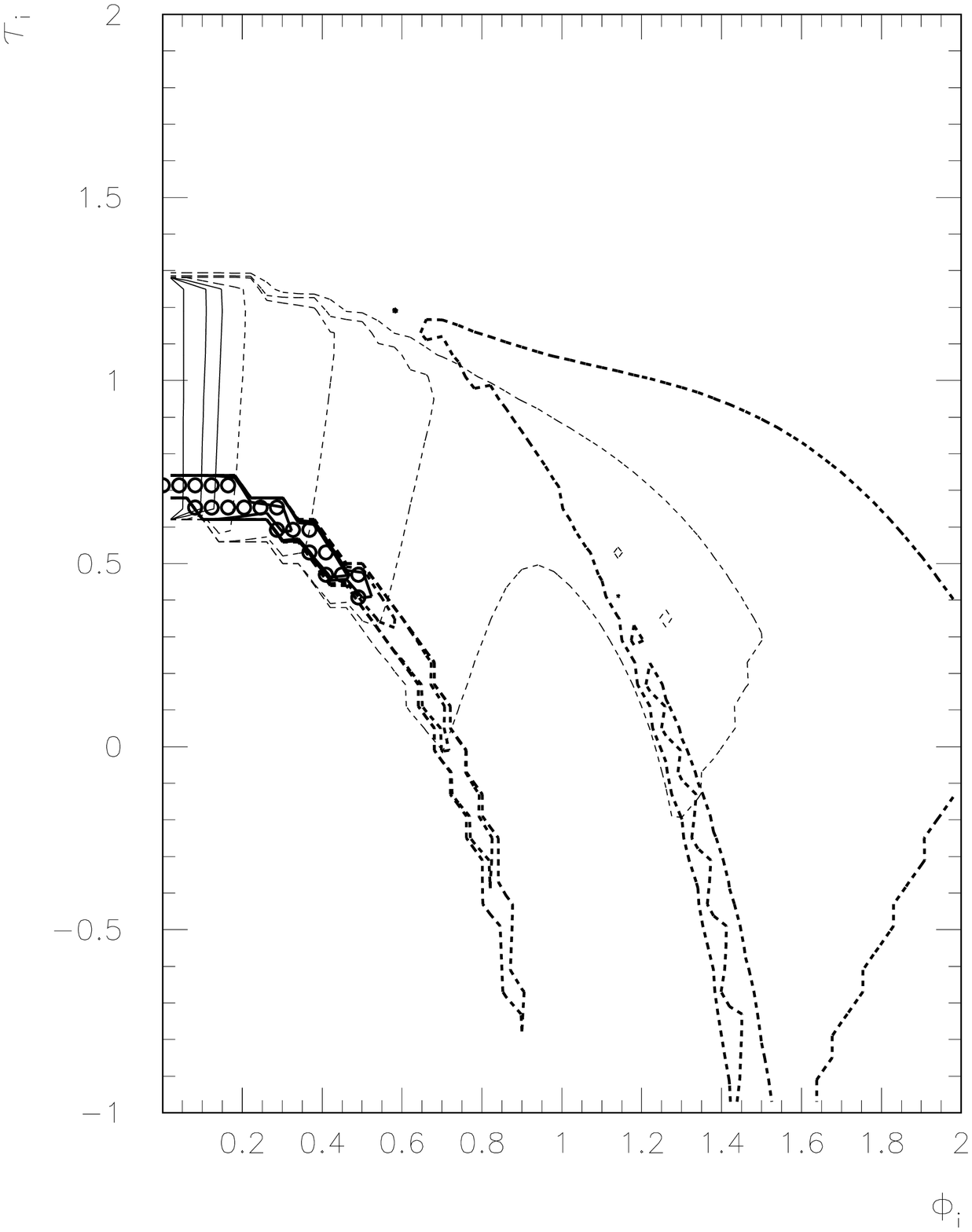}
\includegraphics*[height=7cm]{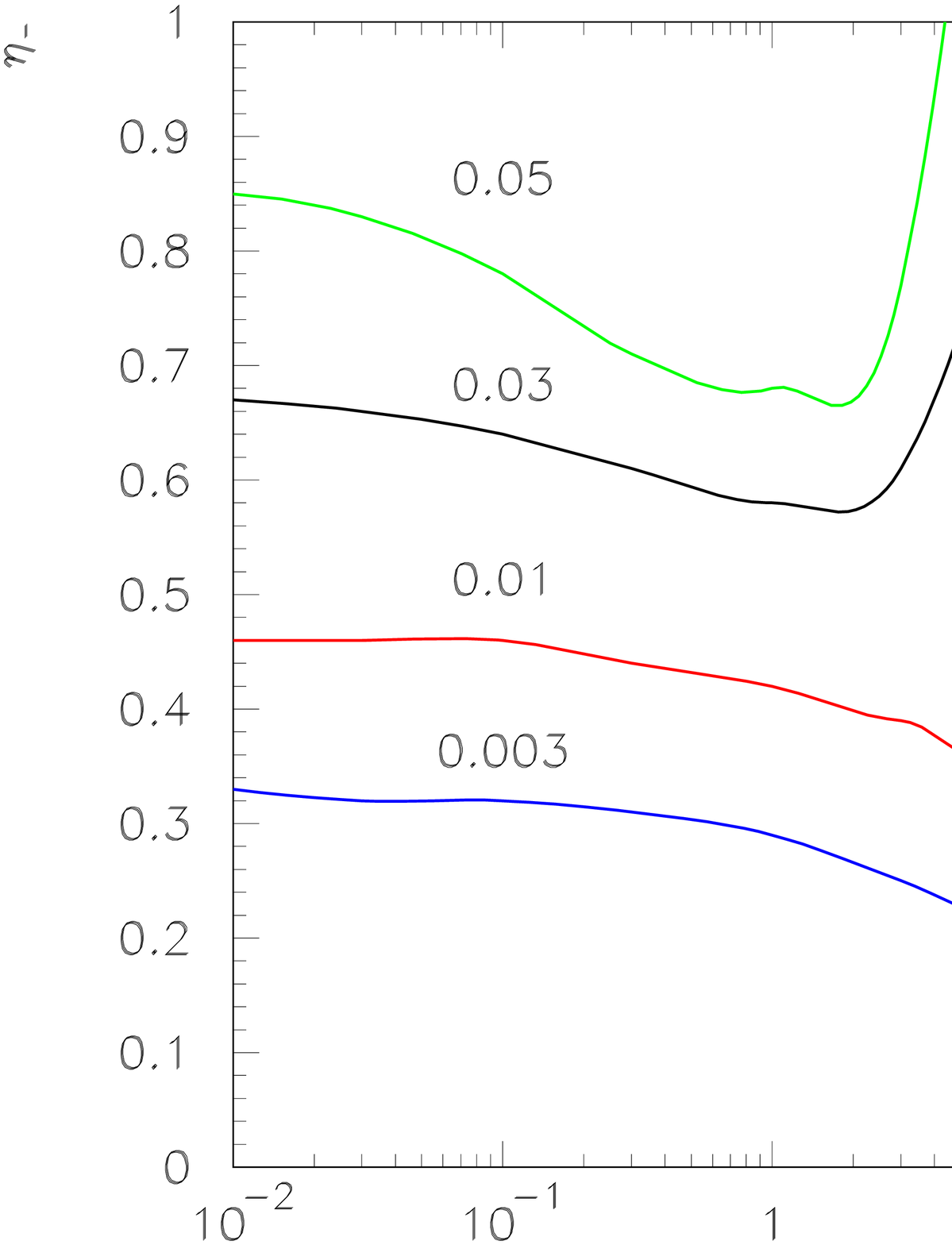}
\end{center}
\caption{\it
Left and centre: Predictions for the number of e-foldings and the spectral index as
functions of the initial conditions for inflation for the Model A (left) 
and Model C (centre). Isocontours of 
$N_\mathrm{ef}=5,10,20,40$ (dashed) and $N_\mathrm{ef}=50,60,100$ (solid) are shown. 
The red circles correspond to 
the spectral index consistent with the WMAP data at 95\% CL (the running of the 
scalar spectral 
index is negligible).
Right: Moduli of negative eigenvalues of $\eta_{ij}$ defined in (\ref{etaij_def})
as a function of $a$ for different values of $1-\Phi_0^2/t_0$.
\label{tab2}}
\end{figure}

{\bf Model A}: The simplest and, in some sense, the cleanest version of D-term stabilisation corresponds to a charged modulus $T$ and a single spectator with 
weight $n=0$. This choice avoids the kinetic mixing between the spectator and the modulus. 
In this case, the scalar D-term potential (\ref{d:vd}) is a function of $\tau$ only and,
if the coupling $g^2$ turns out to be $T$-independent, then the D-terms would approach a constant value 
as $\tau \rightarrow \infty$, thus allowing for a $\tau$-driven inflation of the chaotic type
even without matter fields. Alternatively, one may utilise (\ref{d:vd}) to stabilize
$\tau$, and the choice of the superpotential $W=\frac{1}{2}M\Phi^2$ leads then
to a scalar F-term potential of the form (\ref{fix:vother}).
Note that, in stringy models, the potential stabilising $t$ should vanish in the decompactification limit $t\to\infty$. In the context of (\ref{d:vd}) with $n=0$, this is possible when $g^2$
depends on $t$ in such a way that $g^2\to0$ when $t\to\infty$. As a result, the full scalar
potential can be written as:
\begin{equation}
\label{model_a}
V(\tau,|\Phi|) = \frac{M^2}{6}e^{-2\tau}|\Phi|^2+\frac{\tilde g^2}{2}\left( e^{-\tau}-e^{-\tau_0}\right)^2 ,
\end{equation}
where $e^{-\tau_0}=|Q_X||X|^2/3\delta$ and we have chosen $\tilde g^2=9\delta^2g^2/(e^\tau/2+|\Phi^2|)$ as a $T$-independent rescaled coupling.
The surface plot of the potential (\ref{model_a}) generically exhbits a valley in the
$|\Phi|$ direction. More detailed predictions of this model are presented 
in Figure \ref{tab2}, where we set $\tau_0=\ln2$ and $M^2=10^{-4}\tilde g^2$.

{\bf Model B}: Another simple possibility consists of using a charged $T$ 
field and a single spectator field with
the weight $n>0$. 
One might wish to stabilise $T$ by the D-term potential and to use the superpotential (\ref{fix:super2}) to build a model of new inflation, as outlined in Section~\ref{sec:Tfixed}. 
However, the presence of $|\Phi|$ in the D-term potential generically 
leads to a large contributions to $\eta_\Phi$, which can be suppressed only if the
second term in (\ref{d:vd}) is much smaller than the first term for $t$ at its minimum.
In other words, there would be a triple fine-tuning to achieve a small inflaton mass:
between terms from $V_D$, from $e^K$ and $\partial_\Phi W$. Since all these terms
have in principle different origins, we discard this scenario as relatively improbable.

{\bf Model C}: At this point one may note that, even if  the modulus $T$ is neutral, two spectator fields $X_1$ and $X_2$ with different 
$n_1$ and $n_2$ and charges of opposite signs can stabilise $t$. Thus D-term stabilization looks like a generic phenomenon in a wide class of models. In the following, we focus on this simple case.
For simplicity, we parametrise the supersymmetric D-term potential by 
\begin{equation}
\label{model_c}
V_D = V_0 \left(\frac{t_0}{t}\right)^{2n_1}\left( 1 - \left(\frac{t_0}{t}\right)^\beta \right)^2 \, , 
\end{equation}
where we have taken
\beq
\beta=n_2-n_1>0, \qquad t_0=\frac{1}{2}\left(-\frac{Q_{X_2}|X_2|^2}{Q_{X_1}|X_1|^2}\right)^\frac{1}{n_2-n_1}, 
\qquad V_0=\frac{g^2|Q_{X_1}|^2|X_1|^2}{2^{2n_1+1}t_0^{2n_1}}.
\eeq
There may be some further constraints due to the anomaly cancellation conditions, but there is enough 
freedom in the model to satisfy these conditions without spoiling inflation. For instance, one can easily add more charged 
fields with vanishing expectation values. 
In order to specify completely the form of the scalar potential, we have
to choose the superpotential $W(\Phi)$. Since $V_D$ has been devised primarily to
stabilise $t$, following the considerations in Section \ref{sec:Tfixed}, we shall choose
$W$ as in (\ref{fix:super2}). Given that $V_0\gg \mu^4$, the stabilisation of $t$
is insensitive to the presence of the inflaton potential and the effective potential
for the canonically normalised inflaton field is of the form (\ref{pro:pot2}).

We note that inflation from the $t$ field
is still impossible, since the slow-roll parameters $\epsilon_T$ and $\eta_T$ cannot
be small simultaneously. More precisely, $V_D$ has a local maximum corresponding to
$\epsilon_T=0$ but $\eta<-4/3$. Away from the local maximum, $\eta_T$ changes sign,
but $\epsilon_T$ grows large.
Similarly, with the coupling being $T$-dependent, $T$-driven inflation 
becomes difficult due to the $\eta$ problem that cannot be avoided in this case. 

Characteristic predictions of the potential (\ref{model_c}) for inflationary observables
are depicted in Figure \ref{tab2}, where we set $n_1=\beta=1$, $t_0=1$ and $\Phi_0^2=0.99$.
We also chose $V_0=\mu^4$ (thin lines) to make some features of the potential better visible.
The upper part of the regions of initial conditions with a large number of e-foldings 
corresponds to fields rolling to infinity in the runaway direction (decompactification limit),
while the lower part corresponds to the field evolution to the minimum of the potential
at finite $(\tau_0,\Phi_0)$.
Increasing the D-term contribution, {\em e.g.}~changing $V_0$ to $10\mu^4$
(thick lines), raises the barrier
between the valley and the runaway-roll-down region and the allowed region of the 
initial conditions reduces to a narrow strip corresponding to a single-field inflation described by the potential (\ref{fix:new1}). 

\section{Modulus Stabilisation through the F-Terms}

It follows from the discussion in Section \ref{sec:Tfixed} that inflation
from matter fields is, in principle, possible if one assumes that $T$ 
or $t - |\Phi|^2$ is
fixed. In this Section, we present two scenarios of $T$ stabilisation
through F-terms and discuss whether the presence of these additional terms
in the scalar potential can be compatible with inflation.

We note in advance that, in general, two tunings are necessary: one to
obtain inflation with the correct properties, and the second is to lift
the vacuum energy to make the inflationary energy density positive and
sufficiently large, and to end in the flat space after inflation. The
big question is: to what extent are these tunings independent? If the
cancellation of the vacuum energy is to be supersymmetric, then it is
rather obvious that in general the two tunings must be correlated, as the
$F^2$ terms and the $-3 e^K |W|^2$ term in the effective potential both
depend on the same superpotential $W$.

In models of flux compactification,~\cite{Derendinger:2005ph},~\cite{Blumenhagen:2003vr},~\cite{Giddings:2001yu},
one may obtain the following 
superpotential for the modulus field:
\beq
W_1(T) = \frac{1}{\sqrt{6}}\left(Ae^{-aT}+B\right) .
\eeq
For $B/A<0$, the resulting scalar potential has a minimum for a finite value 
of $t$ in a single-field scenario. However, this minimum corresponds to a 
negative value of the vacuum energy, which would result in an anti-de-Sitter 
Universe after inflation. This problem cannot be circumvented by adding a 
simple, monomial superpotential for the matter field, $W_2(\Phi)\propto 
\Phi^n$, $n=1,2,3$. In the following, we choose $W_2(\Phi)=\tilde A\Phi^3$, 
which has the advantage that the contributions from $W_2$ to the scalar 
potential decouple from those coming from an arbitrary $W_1(T)$, thereby 
ensuring that the resulting scalar potential:
\beq 
\label{potential1} 
V(t,|\Phi|)=\frac{1}{(t-|\Phi|^2)^2}\left[ \frac{A^2a}{12}\left( e^{-2at}(1+
\frac{1}{3}at) 
+\frac{B}{A}e^{-at}\cos(at')\right)+\frac{3}{8} \tilde A^2 |\Phi|^4 \right]
\eeq 
is bounded from below.  The slow-roll parameter $\eta_{t'}$ associated
with the imaginary part of $T$ is proportional to $at$ and typically
large. We can, therefore, assume that $t'$ sits already  at its minimum,
$at'=0,2\pi,\ldots$ for $B/A<0$, when inflation starts. The terms
originating from $W_2$ cannot lift the minimum of the scalar potential to
non-negative values, since, due to the presence of the $|\Phi|^2$ term in
the prefactor of (\ref{potential1}), the minimum of the one-dimensional
potential $V(t,0)$ is a saddle point of the full scalar potential. The
values $\tro,\Phi_0$ of the fields at the minimum are given by the
relations
\begin{equation}
\label{v01}
\frac{3}{8}\tilde A^2 = \frac{A^2a^2e^{-2a\tro}}{36\Phi^2_0} \qquad 
\textrm{and} \qquad
\frac{B}{A} = - e^{-a\tro}\left(1+\frac{2}{3}a\tro\right) ,
\end{equation}
and the negative value of the potential for this field configuration is
\beq
V(\tro,\Phi_0) = \frac{A^2a^2}{36(\tro-\Phi^2_0)}e^{-2a\tro} \equiv -V_0 .
\label{v03}
\eeq
At this level, there is no potential for $\theta=\mathrm{arg}\,\Phi$,
which supposedly arises at the loop level through Yukawa interactions of
$\Phi$. 

If the necessary uplifting does not destroy the saddle point at $(\tro,0)$,
one could impose
the initial conditions for inflation in its vicinity.
The value of $t$ would not then change significantly,
since this field is already at its minimum and this model would realise
single-field inflation.

A possibility for uplifting a negative minimum of the scalar potential
consists in adding a brane-antibrane potential, as arises naturally in flux
compactification models,~\cite{Kachru:2003aw}.
In this scenario, we replace
$V_0$ by $V_0\left(\frac{t_0}{t}\right)^{2n}$.
The relations (\ref{v01}) and (\ref{v03}) then change to:
\begin{equation}
\label{v04}
\frac{3}{8}\tilde A^2 = \frac{A^2a^2e^{-2a\tro}}{36\Phi^2_0(1-\alpha)}, \;
\frac{B}{A} = - e^{-a\tro}\left(1+\frac{2-\alpha}{3(1-\alpha)}a\tro\right) \;\textrm{and}\;
V_0 = \frac{A^2a^2e^{-2a\tro}}{36(\tro-\Phi^2_0)(1-\alpha)} ,
\end{equation}
where $\alpha=\frac{2n}{2+at_0}(1-\Phi_0^2/t_0)$. However, utilising the brane-antibrane
potential typically shifts the saddle point away from $\Phi=0$ and the trajectory
linking the saddle point with the minimum of the potential involves a substantial
change in $t$, which spoils inflation, as the curvature of the potential is typically large in the $T$ direction. This statement is illustrated in Figure \ref{tab2}, where we present the values of the negative eigenvalue $\eta_-$ of the slow-roll matrix:
\begin{equation}
\label{etaij_def}
\eta_{ij} = \frac{1}{2V} \frac{\partial{}}{\partial\phi_i} K^{-1}_{jk} \frac{\partial V}{\partial\phi_k}
\end{equation} 
calculated at the saddle point for $n=1$ and for various values of $a$ and $1-\frac{\Phi_0^2}{t_0}$.
All the negative eigenvalues are of the order of unity and generally decrease with $1-\frac{\Phi_0^2}{t_0}$, but one would need an extreme fine-tuning to get successful
inflation.

\section{Summary and Outlook}

We have investigated in this paper several possible embeddings of simple inflationary potentials  
in supergravity theories, in manners consistent with the recent WMAP3 data. We have concentrated on models with the no-scale structure K\"ahler potential in the inflaton sector, so as
to avoid the $\eta$ problem and to take the advantage of the 
semipositivity of the potential which holds in the exact no-scale structure limit of the models that  we have considered. 
We have found that the no-scale matter sector may lead to acceptable inflation, however  the modulus $T$ must be stabilized 
by some additional mechanism which does not spoil the no-scale structure too violently, and some
apparent fine-tuning is necessary. 

We have considered  D-type stabilisation which in the leading approximation keeps intact  the semi-positivity of the potential, and F-type stabilisation, which breaks this feature. 
With D-stabilisation, the model building becomes unexpectedly  simple, as the scales in the $\Phi$-sector 
(by $\Phi$ we denote the matter-type inflaton) and in the $T$-modulus sector decouple. Depending on the assignments of charges and 
the specific form of the K\"ahler potential for charged fields, both chaotic and symmetry-breaking inflation are possible in this case. 
In particular, slow-roll inflation 
starting from the saddle points in the domain of sub- or near-Planckian field strengths 
With F--stabilisation, the post--inflationary minimum is naturally of the AdS type but, after suitable uplifting 
of the vacuum energy, the structure necessary for  an inflationary epoch may appear. However, with a brane-antibrane-type potential 
for the modulus, the requirement that the uplifted vacuum energy is zero curves the potential too strongly to allow for inflation, at least 
if one stays with a reasonable level of fine tuning between expectation values of the modulus and the matter scalar, say up to one part per mille. 
In fact, in all the cases when $\Phi$ inflation takes place, an interesting fine-tuning between the parameters of the superpotential 
for the $\Phi$ field 
and the parameters of the sector which stabilises the $T$ modulus is necessary. 

In summary, we have demonstrated that inflation from matter fields is a nontrivial feature of supergravity 
models, not only because of the $\eta$ problem, but also because  it is 
subtly intertwined with the issue of moduli stabilisation. 
Nevertheless, models with inflation
driven by the matter sector may be constructed in no-scale supergravity, if the moduli are
 stabilised by some higher-scale dynamics and at the expense of a certain 
level of fine-tuning.

\bigskip

\centerline{\Large \bf Acknowledgements}

\vspace*{0.5cm}

The authors thank D. Langlois and V. Mukhanov for helpful discussions. The visit of D.L. and V.M. to 
ITPUW was supported by the TOK Project  MTKD-CT-2005-029466 ``Particle Physics and Cosmology: the Interface''. Z.L. thanks the CERN Theory Division for hospitality.
This work was partially supported by the EC 6th Framework
Programme MRTN-CT-2004-503369, by the Polish State Committee for Scientific
Research grant KBN 1 P03D 014 26 and by POLONIUM 2005.
K.T.\ acknowledges support from the Foundation for Polish Science 
(Programme START).

\end{document}